\documentclass[aps,prbr,twocolumn]{revtex4-1}

\usepackage[american]{babel}
\usepackage[T1]{fontenc}
\usepackage[utf8]{inputenc}
\usepackage{amsmath}
\usepackage{amssymb}
\usepackage{graphicx}
\usepackage{bm}
\usepackage{color}
\usepackage[pdfborder={0 0 0}]{hyperref}
\usepackage{scrpage2}
\usepackage[absolute]{textpos}
\usepackage[normalem]{ulem}
\usepackage{booktabs}

\newcommand{\ve}{\mathbf{e}}
\newcommand{\veh}{\hat{\mathbf{e}}}
\newcommand{\nvr}{\mathbf{r}}
\newcommand{\vj}{\mathbf{j}}
\newcommand{\vE}{\mathbf{E}}
\newcommand{\vJ}{\mathbf{J}}
\newcommand{\dve}{\delta\mathbf{e}}
\newcommand{\tve}{\tilde{\mathbf{e}}}
\newcommand{\dvj}{\delta\mathbf{j}}
\newcommand{\bs}{\Sigma}
\newcommand{\ts}{\tilde{\sigma}}
\newcommand{\sd}{\sigma_d}
\newcommand{\sh}{\sigma_H}
\newcommand{\dvn}{\delta n(\mathbf{r},t)}
\newcommand{\sD}{\sigma_D}

\allowdisplaybreaks

\begin{document}

\title{Linear Response of Zero-Resistance States}

\author{Maxim Breitkreiz}
\email{breitkreiz@lorentz.leidenuniv.nl}
\affiliation{Instituut-Lorentz, Universiteit Leiden, P.O. Box 9506, 2300 RA Leiden, 
The Netherlands}

\date{April 11, 2017}

\begin{abstract}
A two-dimensional electron system in the presence of a magnetic field and microwave 
irradiation can undergo a phase transition towards a zero-resistance state. 
A widely used model predicts the zero-resistance state to be a domain state, 
which responds to applied dc voltages or dc currents by slightly changing the domain 
structure. Here we propose an alternative response scenario, according to which the domain 
pattern remains unchanged. Surprisingly, a fixed domain pattern does not destroy
zero resistance, provided that the resistance is direction independent. Otherwise, 
if the symmetry of the domain pattern allows a direction dependence of the resistance, 
the domain state can be dissipative. 
We give examples for both situations and simulate the response behavior numerically.
\end{abstract}


\maketitle

\section{Introduction}

At the beginning of this century,
Mani \textit{et al}.\ \cite{Mani2002} and Zudov \textit{et al}.\ \cite{Zudov2003} discovered a new dissipationless state of a 2D electron gas that is exposed to microwave irradiation and an out-of-plane magnetic field \cite{Dmitriev2012}. Upon entering this so-called zero-resistance
state (ZRS), the longitudinal conductivity of the sample drops to zero, while the Hall 
conductivity, unlike in the quantum Hall effect, does not show any 
discontinuity. 
Great experimental and theoretical efforts have been made to understand this
phenomenon, yielding at present strikingly different explanations. 
While theories involving pondermotive forces near the 
contacts \cite{Mikhailov2011} or the effect of radiation on edge states
\cite{Chepelianskii2009} seem to be less likely in view of recent 
measurements \cite{Herrmann2016}, other theories,
predicting the ZRS to be either homogeneous or inhomogeneous,
constitute competing alternatives. 
The radiation-driven electron orbit model combines semiclassics  
with an exact solution of a quantum-harmonic-oscillator problem and explains the ZRS 
in terms wave-packet dynamics and Pauli exclusion principle \cite{Inarrea2005}. According to this
theory the ZRS is homogeneous. 

An in turn different group of theoretical models instead predicts that the ZRS is an 
inhomogeneous domain state.
Here the basic mechanism can been understood via a combination of microscopic calculations 
of the non-equilibrium state \cite{Durst2003, Vavilov2004, Dmitriev2005, Beltukov2016}
 and considerations of the electrodynamics of the system \cite{Andreev2003, Auerbach2005, Finkler2006, Finkler2009}. 
 
This work is based on the domain-state model, which we will now introduce in more detail. 
Microscopic calculations show that the interplay of photon absorption and scattering of the electrons lead to a longitudinal conductivity that oscillates upon changing the microwave frequency, the magnetic field strength, or the electric field strength. For fixed frequency and magnetic field, the conductivity tensor, i.e., the tensor that relates the local electric field $\ve(\nvr)$ and the local current 
density, $\vj(\nvr)=\sigma(e)\,\ve(\nvr)$, can be approximated as \cite{Dmitriev2013}
\begin{equation}
\sigma(e) =\left( \begin{array}{cc}
\sd(e ) & \sh  \\
-\sh & \sd(e )  \end{array} \right),
\label{sigma}
\end{equation}
where the dissipative part $\sigma_d$ depends on the absolute value of the local electric field, 
$e\equiv|\ve(\nvr)|$.
In the parameter range of the ZRS, the dissipative part is negative at $e=0$ and becomes 
positive only above a critical value 
$e_0$, with $\sigma_d'(e_0)\equiv d\sigma_d(e_0)/de>0$.
 This critical field strength is set by the radiation field and is thus typically 
much larger than an external dc field of a
linear-response measurement \cite{Vavilov2004}.

\begin{figure}[b]
\includegraphics[width=\columnwidth]{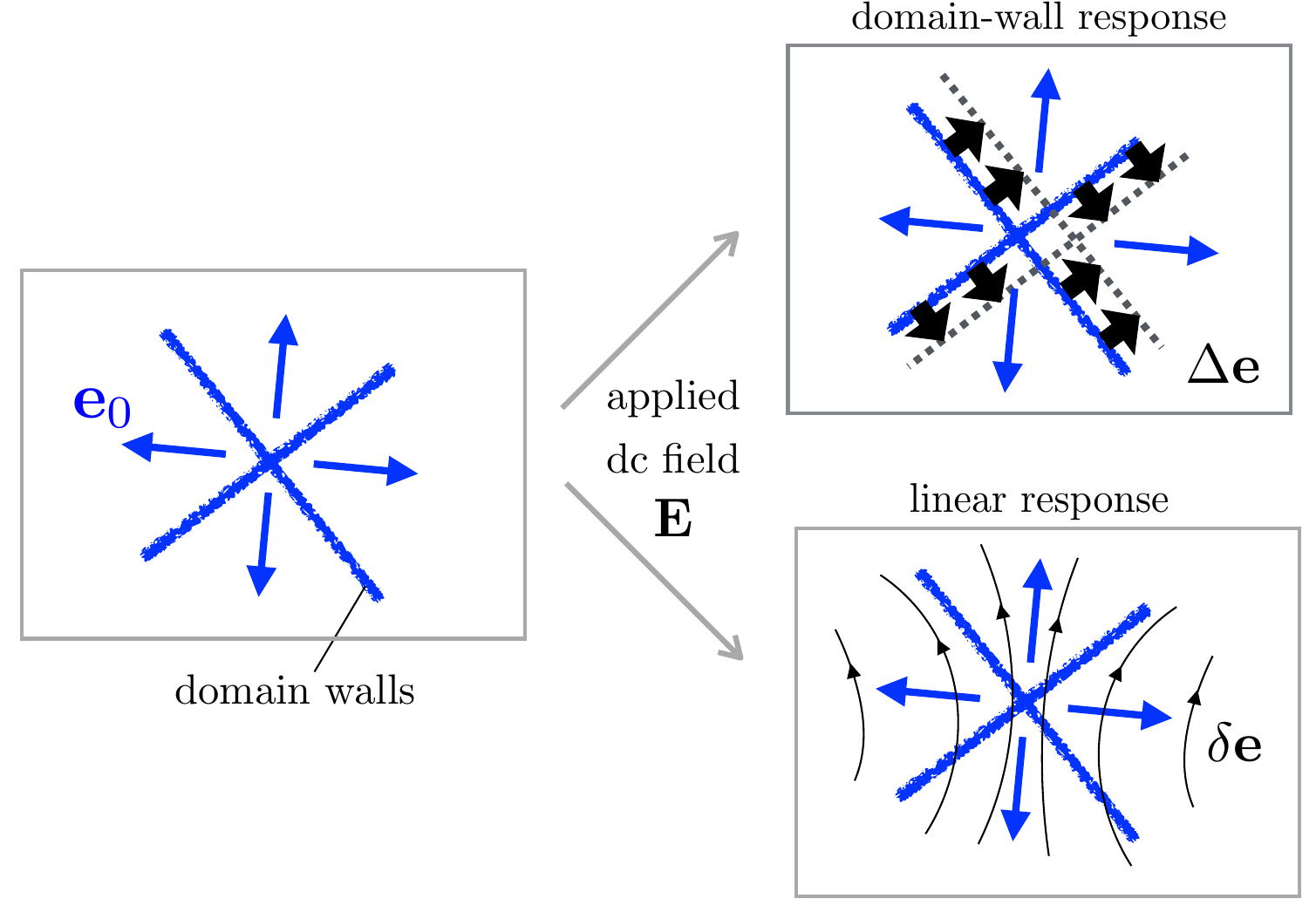}
\caption{Left: Sketch of a general domain realization of the ZRS characterized by 
large internal electric fields $\ve(\nvr)=\ve_0(\nvr)$.
The thick lines indicate the domain boundaries with accumulated charge.
Right: Two possible ways to respond to an external electric field $\vE$. In the 
domain-wall scenario the domain walls shift, leading to large local field changes $\Delta\ve$
with $|\Delta \ve|\sim e_0$. In the linear scenario the domain walls retain their position
and only the low-field pattern $\dve$ with $|\dve|\ll e_0$ changes.}
\label{fig1}
\end{figure}

Naively the theoretical prediction of negative  
$\sigma_d(0)$ seems to imply a negative-resistance state
instead of a ZRS. This however is only true if the system is assumed to be
homogeneous and the effect of boundaries and 
contacts can be neglected. Indeed, numerical simulations of 
a system with boundaries, fixed homogeneous charge distribution, and
negative $\sigma_d(0)$ 
predict the resistance to have positive dissipative part and
a sign-reversed Hall part \cite{Mani2013}, which, however, still 
contradicts experimental observations. 

A different ansatz is to allow for an inhomogeneous charge distribution. 
In this case, neglecting boundary and contact effects, a conventional linear-response
experiment measures the \emph{effective 
conductivity} $\Sigma$, which determines the linear relation
\begin{equation}
\vJ=\bs\,\vE
\label{ohmef}
\end{equation}
between the spatially averaged electric field $\vE=\langle\ve(\nvr)\rangle$ and 
current density $\vJ=\langle\vj(\nvr)\rangle$,
where $\langle\dots\rangle\equiv\int d^2r\,\dots/V$ denotes the spatial average over 
the sample volume $V$. Generally, in inhomogeneous systems the local conductivity 
$\sigma$ does not coincide with the effective conductivity $\Sigma$.
Inhomogeneities in turn can be stabilized if the local conductivity is not positive semidefinite 
\cite{Zakharov1960}, which is the case in the regime of the ZRS for $e<e_0$. 
Andreev \textit{et al}.\ \cite{Andreev2003} thus proposed 
that upon entering the ZRS, the system undergoes a dynamical phase transition
towards a state with an inhomogeneous internal electric field 
\begin{equation}
\ve(\nvr)=\ve_0(\nvr),
\label{sc}
\end{equation} 
which direction can vary in space but the magnitude is fixed to $e_0$ everywhere, 
barring isolated singular points and lines. 
For the average field to vanish, the system must form domains
 \cite{Andreev2003, Finkler2009, Auerbach2005, Dmitriev2013}
with accumulated charge at domain boundaries, as sketched in 
Fig.\ \ref{fig1}. It has been shown that time-dependent fluctuations 
around $\ve_0(\nvr)$ do not diverge with time, signifying the stability of a 
steady domain state characterized by (\ref{sc}) \cite{Andreev2003, Auerbach2005}.

The restriction (\ref{sc}) allows for a variety of possible field patterns $\ve_0(\nvr)$ 
that can be formed upon entering the ZRS regime. While in clean systems the 
system tends to minimize the total length of domain boundaries, impurities
can make the domain pattern more 
complex and disordered \cite{Finkler2006, Auerbach2005}. Measurements that are sensitive to 
local field changes provide experimental support for the domain-state model 
\cite{Dorozhkin2011, Dorozhkin2015} and indicate that the pattern tends to be 
rather complex \cite{Dorozhkin2016a}. An unambiguous evidence for 
the existence of domains, however, is still missing and the exact 
shape and size, could not be observed so far. 

In the following we assume that the system is in a domain state 
with an arbitrary domain pattern and focus on the 
response of the domain state 
to an external homogeneous dc  electric field $\vE$
or an imposed current density $\vJ$. 
Assuming that each domain state must strictly satisfy (\ref{sc}), one can 
 obtain the effective conductivity by averaging the microscopic relation 
$\vj(\nvr)=\sigma(e_0)\,\ve_0(\nvr)$ and comparing with 
(\ref{ohmef}), giving
\begin{equation}
\bs=\sigma(e_0)=\left( \begin{array}{cc}
0 & \sh  \\
-\sh & 0  \end{array} \right),
\label{bs0}
\end{equation}
which is in agreement with experiments. 
The interpretation of the response mechanism is  then
the following: Switching on an infinitesimal electric field $\vE$, the system responds
 in form of an infinitesimal shift of the domain walls, as sketched in Fig.\ \ref{fig1}.  
Compared to the initial state $\ve_0(\nvr)$, 
some domains become shrunk, other domains become expanded 
such that the final state $\tve_0(\nvr)=\ve_0(\nvr)+\Delta\ve(\nvr)$ satisfies the new boundary 
condition, $\langle\tve_0(\nvr)\rangle=\vE$. The induced 
field changes $\Delta\ve(\nvr)$ are zero everywhere except at 
domain boundaries, where they are huge ($\sim e_0$), thus constituting
a locally \emph{nonlinear} response. 

In this work, we argue that the response of the domain state can be locally linear and 
must not involve domain-wall shifting. Our scenario, which we will call
\emph{linear scenario}, involves only small changes of the local electric field $\dve(\nvr)$,
as sketched in Fig.\ \ref{fig1}. In general, this microscopically different mechanism 
results in a different effective conductivity, 
the determination of which turns out to be more difficult then in the domain-wall
scenario. We will derive general symmetry relations that restrict the space of possible
effective-conductivity tensors. These relations fix the effective conductivity to \eqref{bs0}
only if the effective conductivity is isotropic. In the anisotropic case we instead find that the 
domain state can be dissipative.

The outline of this work is as follows. First we define the 
linear-response scenario. In section III we consider the effective conductivity 
in this scenario, and separately discuss the isotropic and the anisotropic cases.
In section IV we discuss the time relaxation of the ZRS that is brought out of the steady state
by a dc voltage for two examples. We conclude in section V.

\section{Linear-response states}

We define a linear-response
state as $\ve(\nvr)=\ve_0(\nvr)+\dve(\nvr)$, composed of a 
high-field pattern $\ve_0(\nvr)$ 
with $|\ve_0(\nvr)|=e_0$ \cite{Andreev2003}, that averages to zero,
and a low-field pattern 
$\dve(\nvr)$ with $|\dve(\nvr)|\ll e_0$ that averages to the external 
electric field to meet the imposed boundary condition $\langle\ve(\nvr)\rangle=\vE$. 
 
The steady high-field pattern $\ve_0(\nvr)$ is rotation free and the corresponding 
current density $\vj_0(\nvr)=\sigma(e_0)\ve_0(\nvr)$ satisfies the stationary
continuity equation. Mathematically, it is thus the solution
of the differential equations
\begin{equation}
\nabla\cdot\vj_0(\nvr)=0,\;\;\; \nabla\times\ve_0(\nvr)=0,
\label{mw1}
\end{equation}
with the boundary condition $\langle\ve_0(\nvr)\rangle=0$. 

Similarly, a steady linear-response state with the current density $\vj(\nvr)=\sigma(e)\,\ve(\nvr)$ 
and the electric field pattern $\ve(\nvr)$ must solve the
same differential equations (\ref{mw1}) with the boundary 
condition $\langle\ve(\nvr)\rangle=\vE$. Due to linearity of the differential 
operators in Eq.\ (\ref{mw1}) and the spatial average $\langle\dots\rangle$,
this is equivalent to the requirement of a steady low-field subsystem
 $\dve(\nvr)$, $\dvj(\nvr)$, which then is the solution of
\begin{equation}
\nabla\cdot\dvj(\nvr)=0,\;\;\; \nabla\times\dve(\nvr)=0,
\label{mw2}
\end{equation}
with the boundary condition $\langle\dve(\nvr)\rangle=\vE$.
The current density of the low-field system is obtained by
subtracting $\vj_0(\nvr)=\sigma(e_0)\,\ve_0(\nvr)$ from $\vj(\nvr)=\sigma(e)\,\ve(\nvr)$
and expanding to linear order in $\dve(\nvr)$ giving
\begin{equation}
\dvj(\nvr)=\ts(\nvr)\,\dve(\nvr),
\label{ohm}
\end{equation}
with the local conductivity 
\begin{align}
\ts(\nvr) =& \sigma(e_0)+\sD\,\veh_0(\nvr)\otimes\veh_0(\nvr) \label{ts1}\\
=& \begin{pmatrix}
0 & \sh \\ -\sh & 0 \end{pmatrix}\nonumber\\ 
&{}+\sD\begin{pmatrix}
\cos^2\phi(\nvr) & \cos\phi(\nvr)\sin\phi(\nvr) \\\cos\phi(\nvr)\sin\phi(\nvr)
 & \sin^2\phi(\nvr) \end{pmatrix},
\label{ts}
\end{align}
where $ \sD\equiv \sd'(e_0)\,e_0 $ and
$\veh_0(\nvr)=\ve_0(\nvr)/e_0=(\cos\phi(\nvr),\sin\phi(\nvr))$ is the direction of 
the high-field electric field, parametrized by the polar angle $\phi(\nvr)$.

The key observation is that the
electrodynamics of the low-field subsystem resembles the electrodynamics 
of a conventional inhomogeneous 
conductor \cite{Dykhne1997}, with an $\nvr$-dependent conductivity  $\ts(\nvr)$, a local electric field
$\dve(\nvr)$, and a local current density $\dvj(\nvr)$ that are induced by the external
electric field $\vE$. 
Like for a conventional conductor, the stability of the steady low-field subsystem 
is thus guaranteed by the positive semidefiniteness 
of the symmetric part of the local conductivity $\ts(\nvr)$.
From (\ref{ts}), this is easily proven to be satisfied for any 
high-field pattern $\ve_0(\nvr)$ for all $\nvr$. A more explicit discussion of the
stability is presented in section \ref{dyn}.

\section{Effective conductivity}
\label{Effcon}

Given that the domain state responds according to the linear scenario
instead of moving the domain walls, we now consider its effective conductivity. 
Particularly interesting
is the question whether the domain state is a ZRS if the system responds according
to the linear scenario, i.e., the field pattern slightly deviates from a pure domain state 
$\ve_0(\nvr)$.

For a given $\ts(\nvr)$ the effective conductivity $\Sigma$ can be defined as
\begin{equation}
\Sigma\;\langle\dve(\nvr)\rangle=\langle\ts(\nvr)\,\dve(\nvr)\rangle
\label{defSigma}
\end{equation}
for all possible $\dve(\nvr)$. From this definition it is in general difficult to 
calculate $\Sigma$ explicitly.
However, thanks to a certain symmetry inherent in the electrodynamics of 2D systems, which
has been found by Dykhne in 1971 \cite{Dykhne1971a, *Dykhne1971}, 
we can derive exact symmetry relations that restrict the space of 
possible tensors $\Sigma$.

We introduce new fields $\dvj'(\nvr)$ and $ \dve'(\nvr)$ via the transformation
\begin{align}
\dvj(\nvr) &= \dvj'(\nvr)+\sh\,R\,\dve'(\nvr) \label{tra} \\
\dve(\nvr) &= 3\,\dve'(\nvr)+\sh^{-1}\,R\,\dvj'(\nvr),
\label{trb}
\end{align}
where $R$ is a $90^\circ$-rotation matrix. Using the 2D-specific relations
\begin{equation}
\nabla\cdot R\,\mathbf{v} = - \nabla\times\mathbf{v}\;\;\;\;\text{and}\;\;\;\;
\nabla\times R\,\mathbf{v} =  \nabla\cdot\mathbf{v},
\end{equation}
one can easily show that the new fields are another solution of (\ref{mw2}) and (\ref{ohm}) 
with the same conductivity tensor (\ref{ts}), like the original fields. 
The two solutions correspond to different boundary conditions, i.e.,
the averaged fields 
$\vJ'=\langle\dvj'(\nvr)\rangle$ and $\vE'=\langle\dve'(\nvr)\rangle$ differ, in general,
 from
$\vJ$ and $\vE$. The effective conductivity, however, does not depend on the fields,
so $\vJ'$ and $\vE'$ must
be related by the same effective conductivity $\bs$ as the fields $\vJ$ and $\vE$. 
Averaging Eqs.\ (\ref{tra})
and (\ref{trb}), and using $\vJ'=\bs\,\vE'$ and $\vJ =\bs\,\vE $, we find 
\begin{equation}
\bs=\Big(1-\sh^{-1}\,\bs\,R\Big)^{-1}\Big(3\,\bs-\sh\,R\Big).
\end{equation}
For a general effective conductivity tensor, this relation is equivalent to
\begin{align}
\det \bs &=\sigma_H^2 \label{sr1},\\
\bs_{12}-\bs_{21} &=2\,\sigma_H, \label{sr2}
\end{align}
where $\bs_{ij}$ are the components of $\bs$. Note that Eqs.\
(\ref{sr1}) and (\ref{sr2}) hold for an arbitrary domain pattern. 
It can be easily seen that a conductivity tensor satisfying 
(\ref{sr1}) and (\ref{sr2}) is positive semidefinite, which allows for 
dissipationless as well as dissipative response.

\subsection{Isotropic effective conductivity}

Isotropy, i.e., direction independence of the effective conductivity
imposes two additional equations,
\begin{equation}
\bs_{11}=\bs_{22}, \; \; \;\;\; \bs_{12}=-\bs_{21}.
\label{iso}
\end{equation} 
Together with the derived symmetry relations (\ref{sr1}) and (\ref{sr2}) this 
fixes the effective conductivity unambiguously to
\begin{equation}
\bs=\left( \begin{array}{cc}
0 & \sh  \\
-\sh  & 0 \end{array} \right).
\label{result}
\end{equation}
This shows that a domain state with an 
isotropic effective conductivity is indeed a ZRS. The linear scenario thus 
correctly reproduces the experiments
 \cite{Mani2002, Zudov2003} in this case.  Isotropy of the effective conductivity
 can be assumed if the domain pattern has 
four-fold rotational ($C_4$) symmetry, or the domain pattern is randomized by 
impurities \cite{Auerbach2005, Finkler2006}.

\subsection{Anisotropic effective conductivity}

\begin{figure}[b]
\includegraphics[width=\columnwidth]{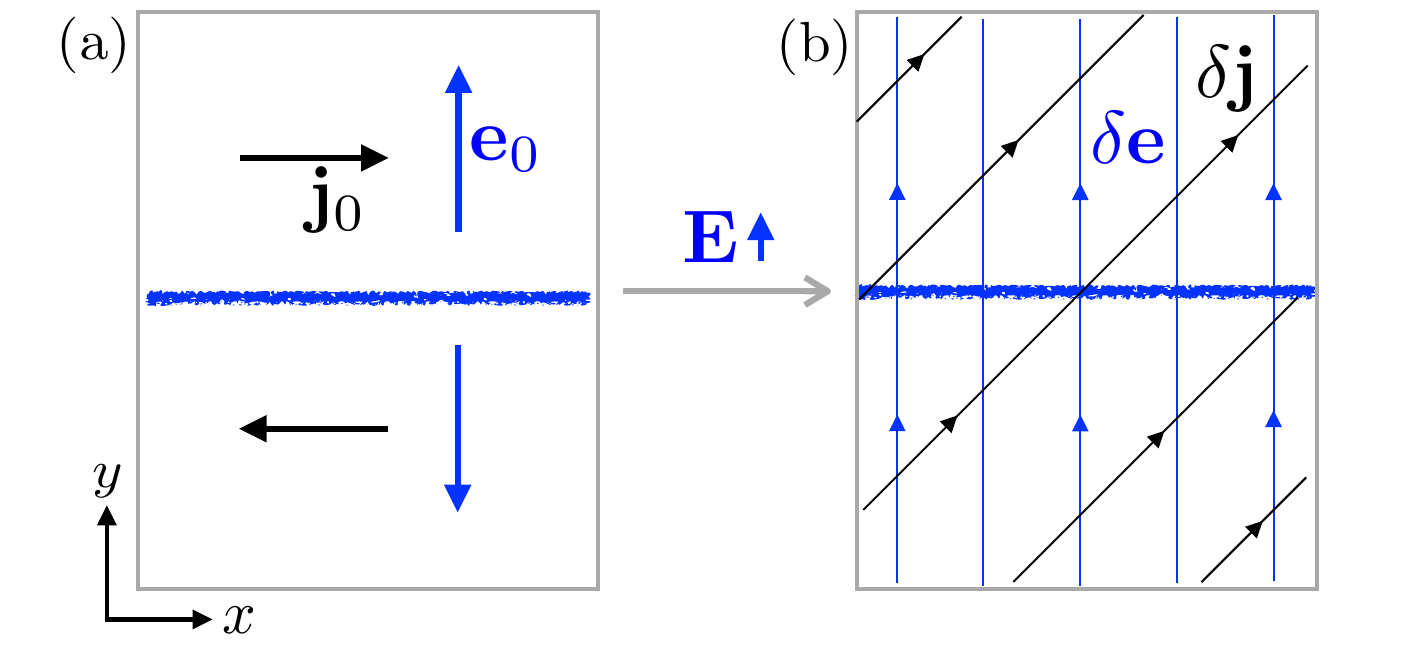}
\caption{Single-domain-wall model (a) The high-field pattern $\ve_0(\nvr)$ 
produced by accumulated charge at the domain wall (thick line) in the middle of the sample. 
(b) Infinitesimal increase of the local electric field by $\dve(\nvr)$ in response to an external 
field $\vE$ applied in the $y$-direction. The effective conductivity is positive definite, hence
the induced current has a component parallel to $\vE$---the response is dissipative.}
\label{fig2}
\end{figure}

Without additional restrictions on the effective conductivity,
Eqs.\  (\ref{sr1}) and (\ref{sr2}) no longer 
guarantee the absence of dissipation. 
In fact, the response can be dissipative in this case, which we show now by 
calculating the effective conductivity for a specific 
 domain pattern. 

We consider a model with a single domain wall separating two
domains with opposite directions of $\ve_0$, as illustrated in Fig.\ \ref{fig2}(a).
Inserting these directions into Eq.\ (\ref{ts}) we find
\begin{equation}
\ts(\nvr)=\begin{pmatrix}
0 & \sh  \\
-\sh  & \sD
\end{pmatrix}.
\label{ex}
\end{equation}
For this simple structure the local conductivity (\ref{ts}) is the same in both 
domains and thus, according to \eqref{defSigma}, equal to the effective conductivity,
\begin{equation}
\Sigma=\begin{pmatrix}
0 & \sh  \\
-\sh  & \sD
\end{pmatrix}.
\end{equation}
Since $\sD>0$ the response is dissipative in contrast to the prediction of
the domain-wall scenario \cite{Andreev2003}.

\section{Dynamical response}\label{dyn}

So far we have discussed the possibility 
of the linear scenario as an alternative to the domain-wall
scenario in the steady regime, i.e., at times when the system had enough time
to rearrange the charge density $\dvn$ after the application of an external field $\vE$.
To decide, which type of response the system will choose, we now 
consider the time dependence of the charge density right after the application of 
an external field.

The dynamics are governed by the continuity equation, the Poisson equation, and 
Ohm's law,
\begin{subequations}
\begin{align}
\frac{d\,n(\nvr ,t)}{dt} &= -\nabla\cdot\vj (\nvr ,t), \label{eqa} \\
\ve (\nvr ,t) &=  -\nabla  Un(\nvr,t)+\vE, \\
\vj (\nvr ,t) &=  \sigma(e(\nvr ,t))\,\ve (\nvr ,t),
\label{eqc}
\end{align}
\label{eqs}
\end{subequations}
where $Un(\nvr,t)$ is the electrostatic potential of the charge distribution, written in terms 
of a positive definite operator $U$, which encodes the Coulomb interaction, acting on the
charge distribution. Considering $U$ as a finite matrix with indices $\nvr$ and $\nvr'$,
its positive definiteness is due to the fact that the diagonal elements are infinite while the
sum over each column or row is finite.

We decompose the charge density into
$n(\nvr,t)=n_0(\nvr)+\dvn$, where $n_0(\nvr)$ is the given charge density of the
accumulated charge at the domain walls that produce the pattern 
$\ve_0(\nvr) =-\nabla \, U\,n_0(\nvr)$
 and $\dvn$ is the time-dependent deviation 
induced by the external field. Then, as previously, the local electric field decomposes 
into the high-field pattern $\ve_0(\nvr)$ and the low-field pattern
$\dve(\nvr,t)= -\nabla \, U\,\dvn+\vE$ and the current density, to linear order in
$\dve$, decomposes into 
$\vj_0(\nvr)=\sigma(e_0)\,\ve_0(\nvr)$ and  $\dvj(\nvr,t)=\ts(\nvr)\,\dve(\nvr,t)$,
where $\ts(\nvr)$ is given in \eqref{ts}. 
The linearization of $\dvj(\nvr,t)$ is valid as long as $|\dve(\nvr,t)|\ll e_0$.

Inserting into Eqs.\ \eqref{eqs},
we find the response entirely in the low-field subsystem, governed by 
\begin{subequations}
\begin{align}
\frac{d\, \dvn}{dt} &= -\nabla\cdot\dvj (\nvr ,t), \\
\dve (\nvr ,t) &=  -\nabla \, U\,\dvn+\vE, \label{dvet}\\
\dvj (\nvr ,t) &=  \ts(\nvr )\,\dve (\nvr ,t).
\end{align}
\label{eqslow}
\end{subequations}
It is useful to consider $\nabla$, $U$, and $\ts_{ij}(\nvr)$ as matrices and 
$\delta e_i (\nvr ,t)$, $\dvn$, $E_i$, and $\delta j_i (\nvr ,t)$ as vectors
by considering the spatial arguments as indices. Doing so and combining
Eqs.\ \eqref{eqslow}, we can write 
\begin{equation}
\frac{d\,\delta n(t)}{dt} = -P\, \delta n(t) -\nabla_{i}\ts_{ij}\,E_j,
\;\;\;\;\;P=\nabla_i^T\ts_{ij}\nabla_j U,
\label{de}
\end{equation}
where the sum over repeated indices $i,j \in \{x,y\}$ is implied and
we used that $\nabla_i$ is skew symmetric. To bring Eq.\ \eqref{de}
in the usual form of linear differential equations, we subtract the 
time-independent part, $\delta\bar{n}(t)=\delta n(t)-\delta n_0(E)$
 with
\begin{equation}
 P\, \delta n_0(E) =\nabla_{i}\ts_{ij}\,E_j.
 \label{sol}
\end{equation}
The solution $\delta\bar{n}(t)=0$ of
\begin{equation}
\frac{d\,\delta \bar{n}(t)}{dt} = -P\, \delta \bar{n}(t)
\end{equation}
is Lyapunov stable if the real parts of the eigenvalues of $P$ are 
non-negative and those that are zero are semi-simple 
(i.e.\ its algebraic and geometric multiplicities coincide).
That this is indeed the case can be seen
by using the Cholesky decomposition $U=LL^T$ to obtain
\begin{equation}
P=\big(L^T\big)^{-1} M L^T, \;\;\;\;\;\;
M= L^T\nabla_i^T\ts_{ij}\nabla_j L .
\end{equation}  
The symmetric part of $M$ can be written as 
\begin{align}
M_S &= \frac{M+M^T}{2}=L^T\nabla_i^T\frac{\ts_{ij}+\ts_{ji}}{2}\nabla_j L \\
& = L^T\nabla_i^T\ts_{S,ij}\nabla_j L,
\end{align}
where $\ts_{S}$ is the symmetric part of $\ts$. Similarly, the skew symmetric 
part $M_A$ is given by the skew symmetric part of $\ts$, which has $\pm\sigma_H$
on its off-diagonal. Since $\sigma_H$ is $\nvr$ independent,
it commutes with $\nabla_i$ and we obtain
\begin{equation}
M_A=\sigma_HL^T\big[\nabla_y,\nabla_x\big]L=0.
\end{equation}
For an arbitrary vector $v$ we can thus write a square form as
\begin{equation}
v^T M v =V^T \ts_S V,
\end{equation}
where $V=(\nabla_xLv,\nabla_yLv)$ is a vector from a squared vector space compared to 
the vector space of $v$.
Since  $\ts_S$ is positive semidefinite in this squared vector space, we conclude that 
$v^T M v \geq 0$ for all $v$, hence $M$ is also positive semidefinite. 
Since $M_A=0$, $M$ is symmetric, consequently its eigenvectors are linearly independent.
These properties are inherited by  
$P$ because it is similar to $M$ and we can conclude that the eigenvalues of 
$P$ are non-negative and those which are zero belong to linearly-independent
eigenvectors, are thus semi-simple. These are 
sufficient criteria for the Lyapunov stability of the steady solution $\delta \bar{n}(t)=0$,
or equivalently $\delta n(t)=\delta n_0(E)$. 

For $\vE=0$, the considerations above are essentially a revision of the arguments 
made in Ref.\ \cite{Andreev2003} on the stability 
of the state $\ve(\nvr)=\ve_0(\nvr)$. For $\vE\neq 0$, however, this 
shows that the linear-response states are also stable, although they deviate
from $\ve(\nvr)=\ve_0(\nvr)$ by $\dve(\nvr)$ with $|\dve(\nvr)|\ll e_0$.

A solution with the boundary condition $\delta n(t=0)=0$ reads 
\begin{equation}
\delta n(t)=\big(1-e^{-P\, t}\big)\,\delta n_0(E),
\label{solx}
\end{equation}
from which we see that only the decaying non-zero modes of $P$ contribute,
which according to \eqref{sol} scale with $E$.
 
Our main conclusion from this is that the application of an external field
$E\ll e_0$ on a domain state with $\ve(\nvr)=\ve_0(\nvr)$ will lead to
small changes of the local electric field, which scale with $E$. In particular,
this implies that the domain walls will not shift, since this would require 
$|\dve_0(\nvr)|\sim e_0$.

We now demonstrate this behavior with two examples: 
The single-domain-wall model from the previous section
and a $C_4$ symmetric model.

\subsection{Single-domain-wall model}

We consider again the single-domain-wall model shown in Fig.\ \ref{fig2}. 
Suppose that for $t<0$ the domain state has the 
high-field pattern $\ve_0(\nvr)$ as shown in Fig.\ \ref{fig2}(a).
At $t=0$ we switch on 
an external electric field $\vE$, so that $\dve(\nvr,t=0)=\vE$,
as illustrated in Fig.\ \ref{fig2}(b).
Since the local conductivity (\ref{ex}) is equal in both domains, the induced current density 
$\dvj(\nvr,t=0)=\ts(\nvr)\,\vE$ is $\nvr$-independent, hence $\nabla\cdot \dvj(\nvr,t=0)=0$.
From this follows immediately $\delta n(\nvr,t)=0$, thus the system will remain in 
the $t=0$ state, which is a linear-response state with $|\dve|=|\vE|\ll e_0$.

For this particular domain pattern, the linear-response state coincides with the
state at the instance when the electric field was switched on.
In general, this is not true as is shown in the next example.

\subsection{Checkerboard model}

\begin{figure}[b]
\includegraphics[width=\columnwidth]{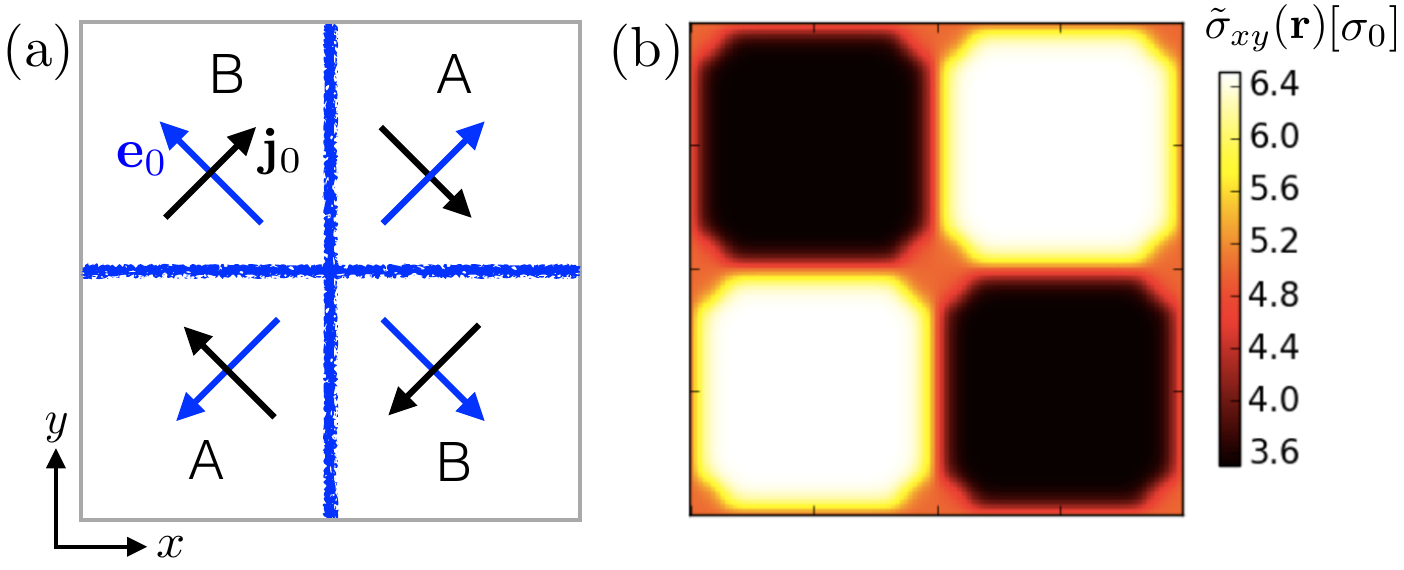}
\caption{Checkerboard model. (a) The domain pattern consisting of 
four domains with different directions of $\ve_0$ and the resulting circulating current $\vj_0$. 
(b) $\ts_{x,y}(\nvr)$ on a $200\times 200$ grid with $\sD=3\,\sigma_0$ and 
$\sh=5\,\sigma_0$. }
\label{fig3}
\end{figure}

The domain pattern for this model is shown in Fig.\ \ref{fig3}(a). According to Eq.\ \eqref{ts}, 
the local conductivity in the domains A and B reads 
\begin{align}
\ts(\nvr)=
\begin{cases}
\begin{pmatrix}
\tfrac{\sD}{2} & \sh+\tfrac{\sD}{2}  \\
-\sh +\tfrac{\sD}{2} & \tfrac{\sD}{2}
\end{pmatrix} & \nvr \in \text{A}  \vspace*{6pt} \\ 
\begin{pmatrix}
\tfrac{\sD}{2} & \sh-\tfrac{\sD}{2}  \\
-\sh -\tfrac{\sD}{2} & \tfrac{\sD}{2}
\end{pmatrix} & \nvr \in \text{B} .
\end{cases}
\label{tssym}
\end{align}
At $t=0$, the local current density has a finite divergence, so the charge density will
evolve, governed by Eq.\ \eqref{de}. We simulate the time evolution numerically by 
discretizing the time with a finite time step $dt$ and
discretizing the space by an $N\times N$ grid with periodic boundary conditions in 
both directions. We measure the length in units of the domain length $l$, so that $\nvr=(x,y)$
 with $x,y=2l\, i/N$, $i \in [0,N-1]$.

To provide a reasonable description on the discretized space,
we have to smoothen the domain boundaries over a few space points, i.e., 
find a continuous version of Eq.\ \eqref{tssym}. To do so,
we convolute the $N\times N$
matrices $e_{0,x}(\nvr)$ and $e_{0,y}(\nvr)$ with a Gaussian kernel of size $0.15\,l$
and standard deviation $0.1\,l$ to determine the continuous version of the 
angle $\phi(\nvr)=\arctan(e_{0,y}(\nvr)/e_{0,x}(\nvr))$ in \eqref{ts}.
At points where the angle is not defined, 
we suppress $\sD$ by the function $1/2+\tanh[(|\nvr|-0.15\,l)/0.05\,l]/2$. 
The resulting spatial dependence of $\ts$ is shown in Fig.\ \ref{fig3}(b).
We checked that the variation of these parameters does not
have qualitative influence on the result.

\begin{figure*}[t]
\includegraphics[width=\textwidth]{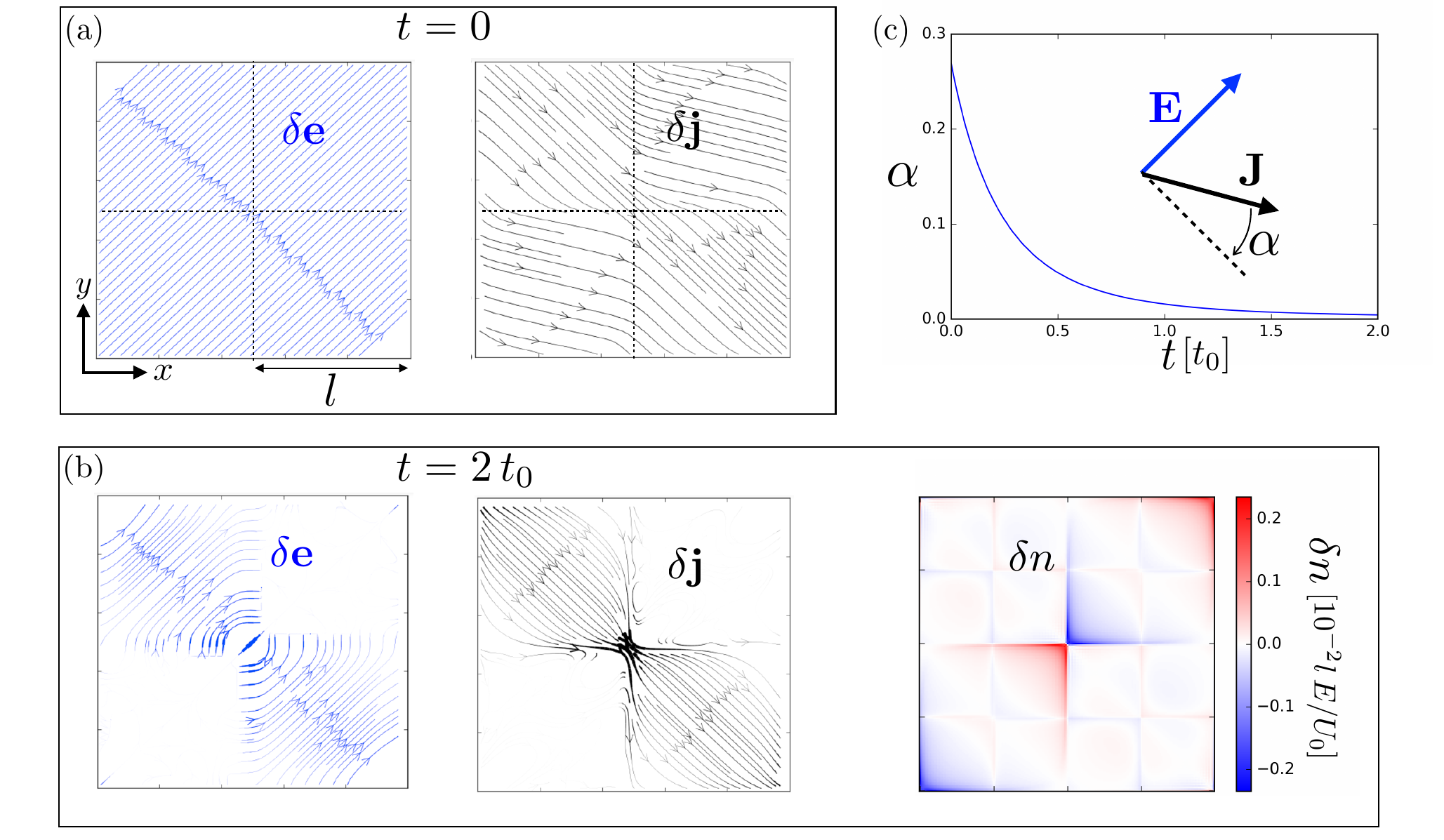}
\caption{Time evolution of the checkerboard model. (a) The initial state at $t=0$,
when an external electric field $\vE=E(1,1)/\sqrt{2}$ has been turned on. (b) 
The state at $t=2\,t_0$, where $t_0=(10^{-2}lE)^2/U_0\sigma_0$. Further time evolution gives no visible changes 
from which we conclude that the system essentially reached the steady state. In
the stream plots the line width is proportional to the field magnitude. All 
magnitudes are proportional to $E$. (c)
Time evolution of the angle between the applied field $\vE$ and the induced 
total current $\vJ(t)=\langle\dvj(\nvr,t)\rangle$. As predicted analytically in Section \ref{Effcon}, 
in the steady state ($t \gtrsim 2\,  t_0$), the averaged current flows perpendicular
and is equal to $\Sigma\, \vE$ with $\Sigma$ form \eqref{efend}. 
The parameters for these plots are $\lambda=l$, $\sD=3\,\sigma_0$, $\sh=5\,\sigma_0$,
and $dt=5\times 10^{-4}t_0$. 
The spatial grid is $200\times 200$, which turns out to be sufficient since already a 
halved precision gives no visible differences in the plots 
(except for a coarser grain).}
\label{fig4}
\end{figure*}

\begin{figure*}[t]
\includegraphics[width=0.9 \textwidth]{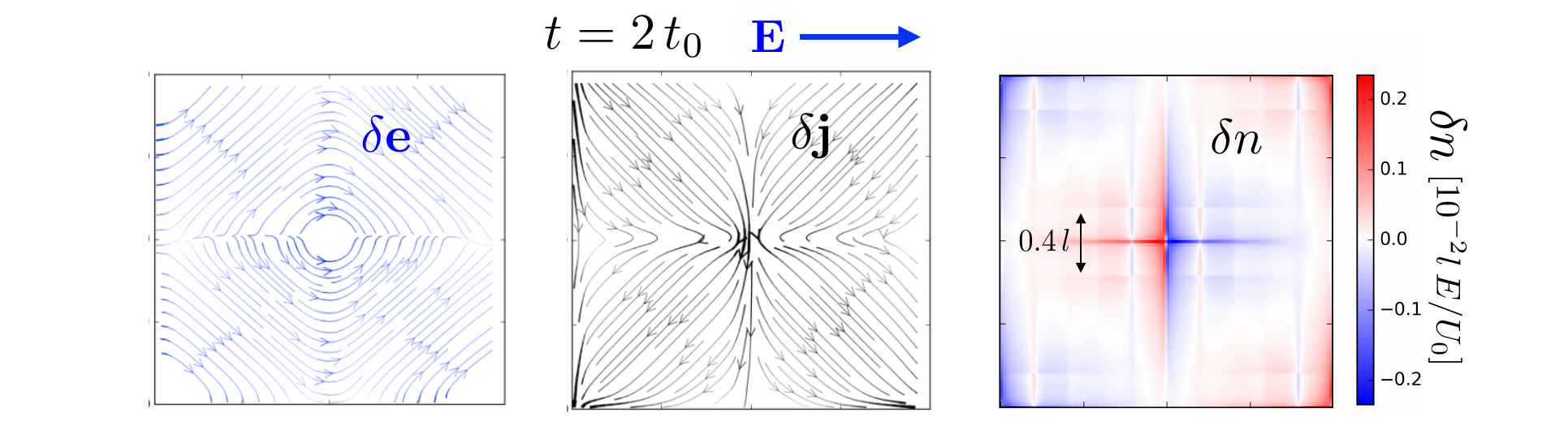}
\caption{Steady state of the checkerboard model for 
$\vE=E(1,0)$ and $\lambda=0.4\, l$. Otherwise same parameters as in Fig.\ \ref{fig4}.
The length scale of
the weak substructure, visible in $\delta n$, corresponds to the size of the interaction
kernel, is thus a numerical artifact, which however has no influence on the effective 
conductivity $\Sigma$.}
\label{fig5}
\end{figure*}

We approximate the action of the interaction operator $U$ on 
$\delta n$ by a convolution of $\delta n$ with a kernel 
\begin{equation}
U_{x,y}=\frac{U_0}{\sqrt{(x^2+y^2)/\xi^2+\eta^2}}
\end{equation}
of size $\lambda \times \lambda$ and with parameters set to $\xi=0.1\,l$ and
$\eta= 10^{-2}$ ($\eta$ can be seen as a finite out-of-plane component and we consider
numerically the limit $\eta\to 0$). 
The variation of these parameters and the size of the 
kernel within the physical
parameter regime (which is restricted by the requirement of a positive definiteness of $U$),
does not lead to qualitative differences. Representative numerical 
results are summarized in Fig.\ \ref{fig4}.

As predicted by our symmetry considerations, in the steady state the average current 
flows perpendicular to the applied field, according to the simple relation
$\vJ=\Sigma\, \vE$ with
\begin{equation}
\Sigma=\begin{pmatrix}
0 & \sh  \\
-\sh  & 0
\end{pmatrix}.
\label{efend}
\end{equation}
In contrast to this, the local fields and the charge density acquire  a non-universal 
structure during the 
evolution, which depends on the details of the interaction,
domain length, and the direction of the applied electric field. This dependence is illustrated  
in Fig.\ \ref{fig5}, where we show plots of the steady state for the electric field $\vE=E(1,0)$ 
and $\lambda=0.4\, l$ (cf.\  Fig.\ \ref{fig4}: $\vE=E (1,1)/\sqrt{2}$ and $\lambda=1\,l$ ). 
Comparing the figures, we see that the local field patterns change dramatically. The effective
conductivity \eqref{efend} stays the same. 

It is interesting to compare this ZRS to the quantum
Hall effect for magnetic fields at the Hall plateaus. 
The Hall effect also shows a purely transversal 
resistance, hence a very similar macroscopic response. The microscopic 
current flow, however, turns out to be different: While in the Hall effect bulk 
states are localized and current is carried entirely by the edge states, 
in the domain state current flows through the bulk, albeit in an 
inhomogenous pattern. 

\section{Conclusion}

In conclusion, we have considered the response of domain states to external fields 
(induced by a dc voltage)
that are much smaller then the internal fields within the domains
(induced by microwave radiation). 
We have proposed a new
response mechanism, which, contrary to the established one, does not involve domain-wall shifting.
In our view, small external fields lead only to small modulations of the local fields, 
leaving the domain patterns unchanged. 
The theoretical justification of the linear scenario is based on the fact that small
deviations from the pure domain state are not unstable, which we have shown
by analyzing the electrodynamics of a general domain state. We tested 
these predictions on two specific realizations numerically.  

Our main results address the effective conductivity of the domain state in the 
linear scenario: 
If the effective conductivity is  
isotropic, which is the case if the domain pattern is chaotic or $C_4$ symmetric, 
then the response is dissipationless. Otherwise
the response can be dissipative, which we have shown for
the single-domain-wall pattern---the energetically most favorable 
pattern in a clean system \cite{Andreev2003}.

Combining this result with the fact that disorder can pin the domain walls 
in a chaotic pattern \cite{Auerbach2005, Finkler2006},
this work supports the idea that the radiation-induced ZRS is a 
disordered domain state, where the disorder is strong enough to
allow for a chaotic domain pattern. Clean domain states, instead, are 
allowed to have a dissipative response. The explicit value of the
longitudinal conductivity is presumably non-universal in this case.
This is in stark contrast to the domain-wall response, which predicts 
strictly zero resistance in the clean case and 
dissipative response in the regime of pinned domain walls 
\cite{Auerbach2005}.

Measurements that are sensitive 
to local changes of the electric 
field \cite{Dorozhkin2015, Dorozhkin2016a} 
indicate that the domain structure is indeed rather complex and show that 
the local electric fields change proportional to the applied voltages.
These observation, hard to reconcile with the domain-wall-response 
scenario, are in qualitative agreement with our theory. 

We stress that this work is based on the assumption that the system is in a
domain state, which at present appears to be the dominant picture for the ZRS but
not indisputably established one. Steadiness 
of the domain pattern is also an essential ingredient for 
our analysis, which is a justified 
assumption in radiation-induced ZRSs \cite{Finkler2009, Dorozhkin2011}. 
Time-dependent patterns, however, may occur in many other cases, e.g., 
including the zero-differential-resistance states in dc biased 2D electron gases in strong
magnetic fields \cite{Bykov2007}, where the domain structure moves between 
boundaries of the sample. The extension of the present analysis to time-dependent 
domain patterns remains a subject for future work. 

\acknowledgments

Discussions with C.\ Timm and F.\ von Oppen are gratefully acknowledged. This work is financed by 
the Netherlands Organisation for Scientific Research (NWO).

\bibliography{library}

\end{document}